\documentclass[prl,twocolumn,preprintnumbers,superscriptaddress,showpacs]{revtex4-1}

\usepackage{graphicx}
\usepackage{bm}
\usepackage{amsmath}
\usepackage{amssymb}
\usepackage{slashed}
\usepackage[colorlinks,pdfstartview=FitH]{hyperref}
\hypersetup{linkcolor=blue,citecolor=blue,filecolor=black,urlcolor=blue}
\newcommand\sect[1]{\emph{#1.}---}

\def \be {\begin{equation} }
\def \ee {\end{equation}}

\def \bem {\begin{multline}}
\def \eem {\end{multline}}

\def \bes {\begin{subequations} }
	\def \ees {\end{subequations}}

\def \lqcd{ \Lambda_{ {\rm QCD} } }
\def \Nc{ N_{ {\rm c} } }

\def \<{\langle}
\def \>{\rangle}
\def \+{\dagger}
\def \({\left(}
\def \){\right)}
\def \[{\left[}
\def \]{\right]}

\usepackage[normalem]{ulem}

\begin{document}
\title{Quark fluctuations and anisotropic confinement in magnetic fields}

\author{Gaoqing Cao}
\affiliation{School of Physics and Astronomy, Sun Yat-Sen University, Guangzhou 510275, China.} 
\author{Toru Kojo}
\affiliation{Key Laboratory of Quark and Lepton Physics (MOE) and Institute of Particle Physics, Central China Normal University, Wuhan 430079, China.}
\date{\today}

\begin{abstract}

We study the string tension $\sigma$ of quantum chromodynamics (QCD) in external magnetic fields $B$ by combining the quasi-particle treatments for quarks with the collective treatments for gluons. We utilize the $1/\Nc$-expansion ($\Nc$: number of colors) and the framework of Wilson's strong coupling expansion at long distance, and compare the resulting string tensions with the lattice data. The effects of magnetic fields are taken into account through the dielectric functions, which are evaluated from the one-loop polarization due to light flavor quarks with $B$-dependent effective masses $M_{\rm f}(B)\ ({\rm f}=u,d,s)$. All Landau levels are adopted in the calculations so that we can explore the string tensions from weak to strong magnetic fields. For the effective quark masses, we use a simple parameterization: $M_{\rm f}(B) = M_{\rm f}(0) + C_B |q_{\rm f}B|^{1/2}$ with $q_{\rm f}$ the electric charge of a given flavor and $C_B$ an adaptive parameter. 
We find that the parameter $C_B$ needs to be small, $C_B \lesssim 0.05$, to qualitatively reproduce the trends of lattice QCD results for $0 \leq |eB| \lesssim 1.5\, {\rm GeV}^2$. The quantitative agreement is found for $C_B \simeq 0$ and the strong coupling $\alpha_S \simeq 5$ in the infrared.
\end{abstract}


\maketitle

\section{Introduction}

The quantum chromodynamics (QCD) continues to reveal its rich structures in various extreme conditions, including a quark-gluon plasma at high temperature and a dense matter at high baryon density \cite{Yagi:2005yb}. In recent years, the main interests of QCD community have been greatly extended to study the QCD matter in strong electromagnetic fields \cite{Deng:2012pc}  or in fast rotation \cite{STAR:2017ckg} which are both realized in the relativistic heavy ion collisions. To mention some, neutral pion superfluidity in parallel electric and magnetic fields~\cite{Cao:2015cka}, the anomalous magnetovorticity effect induced by the interplay between the magnetic and rotational effects~\cite{Hattori:2016njk}, etc. The QCD system with strong magnetic fields may be also relevant to the interior of neutron stars \cite{Olausen:2013bpa,Turolla:2015mwa,Baym:2017whm} and the early universe \cite{Grasso:2000wj}.

In addition to the phenomenological importance, QCD in strong magnetic fields can be regarded as a useful laboratory to test our understanding on the nonperturbative aspects of QCD. The lattice QCD simulations do not have the sign problem for this extreme condition, and the numerical calculations have been performed up to $ |eB| \sim 3\,{\rm GeV^2}$  \cite{Bonati:2014ksa,Bonati:2016kxj}. This energy scale is much larger than the typical QCD scale, $ \lqcd^2 \sim 0.04\,{\rm GeV}^2 $, and is beyond the achievable strength in realistic systems. The lattice simulations \cite{Bali:2011qj,Bali:2012zg} have confirmed the (chiral) {\it magnentic catalysis} at zero temperature, a theoretical prediction that the chiral symmetry breaking is triggered or enhanced by magnetic fields \cite{Gusynin:1994re,Gusynin:1994xp}. Meanwhile, they also found unexpected phenomenon, {\it inverse magnetic catalysis} \cite{Bali:2011qj,Bali:2012zg}, in which the chiral symmetry restoration occurs at a lower temperature for a larger magnetic field. Several scenarios have been proposed to explain this unusual phenomenon \cite{Fukushima:2012kc,Kojo:2012js,Hattori:2015aki,Chao:2013qpa,Cao:2014uva,Mao:2016fha}.
 
The magnetic fields also affect the confining forces between quarks. The static potential between a pair of heavy quark and antiquark has been measured in the lattice QCD simulations \cite{Bonati:2014ksa,Bonati:2016kxj}. For increasing $B$, the confining force in the directions perpendicular to magnetic fields gets slightly stronger, while in the parallel direction it becomes considerably weakened. Furthermore, the authors conjectured that at very large $B$ the string tension vanishes in the parallel direction, which was named {\it deconfinement magnetic catalysis}. The key to understand this behavior should be the dynamics of quarks.  In the directions perpendicular to the magnetic field ${\bf B}=B\hat{x}_3$, their motions are quantized into the Landau levels (LLs); and the quasi-particle energy becomes $E_n^{ {\rm f} } (p_3) = \sqrt{ {\bf p}_3^2 + 2n B_{ {\rm f} } + M_{ {\rm f} } (B)^2} ~(n=0,1, \cdots) $ where $B_{ {\rm f} } = |q_{ {\rm f} } B|$, ${\bf p}_3$ is the momentum along the $B$-direction, and $M_{ {\rm f} }$ is the effective mass of quarks which may change with the magnetic fields. Upon the quantization in the perpendicular directions, each LL has a degeneracy proportional to $B_{ {\rm f} } $. 
Especially , {\it if} $M_{ {\rm f} } (B)$ is not significantly larger than $\lqcd$, the lowest LL (LLL) can accommodate many quarks with low energy $E \sim \lqcd$, as conjectured in \cite{Kojo:2012js,Hattori:2015aki} to explain the lattice result for the chiral condensate~\cite{Bali:2011qj,Bali:2012zg}: $\langle \bar{\psi} \psi \rangle \propto B$ for $B > \lqcd^2$.
The $B$-dependence of $M_{{\rm f}}$ should be important as well in exploring the screening and string breaking effects which are both long-range phenomena.

In this work, we study the anisotropy of confining forces, and, more importantly, the interplay between deconfinement magnetic catalysis and (chiral) magnetic catalysis, paying special attention to the impacts of the chiral effective masses on the string tensions. To compute the string tensions, we use the Wilson's strong coupling expansion, assuming that the coupling constant is strong enough at long distance. On the other hand, the evaluation of their modifications by magnetic fields will be based on the quasi-particle picture for quarks. We take the one-loop polarizations as the representative of such contributions, which can be regarded as small corrections in the limit of large $\Nc$ \cite{tHooft:1973alw}. The polarization effect depends on the $B$-dependence of effective quark masses, so we test several parameterizations for the effective masses. Provided that the effective masses do not grow rapidly with $B$, our simple procedure turns out to well reproduce the $B$-dependent quantities measured on the lattice. The semi-quantitative success of this simple approach gives us a hope to understand the nonperturbative aspect of QCD in other extreme conditions, e.g. large baryon density. 
 
We will use the following conventions in this paper: The notations $``\parallel"$ and $``\perp"$ stand for the directions parallel (along $x_0$ or $x_3$) and perpendicular (along $x_1$ or $x_2$) to the external magnetic fields, respectively. For a four-vector $x_\mu = (x_\parallel, {\bf x}_\perp)$, $x_\parallel^2 = g_\parallel^{\mu \nu} x_\mu^\parallel x_\nu^\parallel= x_0^2-x_3^2$ and $x_\perp^2 =-{\bf x}_\perp^2= g_\perp^{\mu \nu} x_\mu^\perp x_\nu^\perp=-x_1^2-x_2^2$. We often represent positive constants by $c$ or $c'$.

\section{Confinement dynamics}

\subsection{Dielectric functions at one-loop }

We are interested in the color-electric flux tube between two opposite color charges. In energy momentum space, the effective Lagrangian for the color-electric fields in external magnetic fields should take the form:
\begin{eqnarray}
{\cal L}^B_{ {\rm eff} } 
= \frac{\, 1 \,}{\, 2 \,}  \int_k \left(  
\xi^B_{\parallel} (k) \,  ( {\bf E}_{\parallel}^a)^2  + \xi^B_{\perp} (k) \, ( {\bf E}_{\perp}^a)^2  \right) + \cdots,
\end{eqnarray}
where $\int_k = \int {\rm d}^4 k/(2\pi)^4$, $``\cdots"$ includes the color magnetic fields and higher orders of color-fields.
 The coefficients $\xi_\parallel^B$ and $\xi_\perp^B$ are dielectric functions which take into account the magnetic field effects through quark loops.
 In vacuum, $\xi_\parallel^{B=0} (k) = \xi_\perp^{B =0} (k) = 1$. For larger dielectric functions, the color-electric fields cost more energy to be excited thus prefer smaller amplitudes -- this weakens the string tension.

Let us estimate the dielectric functions and their disparity at one-loop level. Thanks to the constraint from the gauge invariant form, we only have to compute the polarization function. There are three components
\be
\Pi^{\mu \nu} = \Pi_{E\parallel} P^{\mu \nu}_{E \parallel} + \Pi_{M\perp} P^{\mu \nu}_{M \perp} +  \Pi_{ {\rm mix} }  P^{\mu \nu}_{ {\rm mix} } \,,
\ee
where $P^{\mu \nu}_{ E\parallel} = g_\parallel^{\mu \nu} - k_{\parallel}^\mu k^\nu_{\parallel} / k^2_{\parallel}$, $P^{\mu \nu}_{M \perp} = g_\perp^{\mu \nu} - k_{\perp}^\mu k^\nu_{\perp} / k^2_{\perp}$, and $P^{\mu \nu}_{ {\rm mix} } = g^{\mu \nu} - k^\mu k^\nu / k^2 - P^{\mu \nu}_{E \parallel}  - P^{\mu \nu}_{ M \perp} $ are projectors. For each component $i= ( E \!\! \parallel,\,  M \!\! \perp,\, {\rm mix})$, 
the effective coupling in the presence of magnetic fields, $(g_i^B)^2$, includes the quark-loop polarization effects and is given by
\be
\frac{g_{ 0 }^2 }{\, 1+ g_{ 0 }^2 ( \Pi^{ B=0 }_{ {\rm A+ c+ q} } + \Pi_{ {\rm q} }^B - \Pi_{ {\rm q} }^{ B=0} )_i\,}
= \frac{g^2(k^2) }{\, 1 +  g^2 (k^2) (\Delta \Pi_{ {\rm q} }^B )_i \,} \,,
\ee
where  $\Pi^{B=0}_{ {\rm A+ c+ q} }$ is the sum of the vacuum contributions from gluons, ghosts, and quarks, $g_0$ is the running coupling at scale $\mu_0$ such that $\Pi^{B=0}_{ {\rm A+ c+ q} } (\mu_0^2) =0$, and $g^2 (k^2) = g_{ 0 }^2 / \left( 1+ g_{ 0 }^2  \Pi^{B=0}_{ {\rm A+ c+ q} } (k^2)  \right)$ is the effective coupling in vacuum. Then, the dielectric functions are given by $\xi_i^B=g^2/ (g_i^B)^2 = 1 + g^2 (\Delta \Pi_{ {\rm q} }^B )_i$, where the effects of magnetic fields are summarized in the UV-finite function $\Delta \Pi^B_{ {\rm q} } = \Pi_{ {\rm q} }^B - \Pi_{ {\rm q} }^{ B=0} $. 
In actual calculations, we assume the vacuum coupling $g^2(k^2)$ to freeze as a fixed constant at some scale around $\mu_0 \sim 1$ GeV \cite{Deur:2016tte}  for the $B$ domain to be explored. 
More details will be discussed below when we compare our results with the lattice data.

The quarks entering the loop have spins parallel or anti-parallel to the magnetic fields and belong to LLs. The interactions with gluons may cause the spin-flip; there are two types of quark-gluon coupling
\begin{eqnarray}
 \int_x \bar{q} \gamma_\parallel t_a q A_a^\parallel\,,~~~~~~~~ \int_x \bar{q} \gamma_\perp t_a q A_a^\perp\,,
\end{eqnarray}
where the first term is the spin conserving interaction, while the latter is the spin flipping one. 

At large $B$, the polarization of the gauge fields $A_\parallel$ at small momenta is dominated by the process in which both quark lines are LLLs. Keeping only the LLL in the loop for each flavor, the form of the polarization function for $A_\parallel$ is
\begin{eqnarray}
 \Pi_{ \mu \nu}^{ {\rm 4D, f} } (k) = \, B_{ {\rm f} } \, {\rm e}^{- {\bf k}_\perp^2 /(2B_{ {\rm f} }) }\,  \Pi_{ \mu \nu}^{ {\rm 2D, f} } (k_\parallel)+ \cdots,
\end{eqnarray}
then the total polarization function is the sum of all the three flavor contributions. There is an overall factor $B_{ {\rm f} }$ associated with the Landau degeneracy, and the exponential $ {\rm e}^{- {\bf k}_\perp^2 /(2B_{ {\rm f} }) } $ is the conversion factor from the plane wave bases to the LLL bases. The left part $ \Pi_{\mu \nu}^{ {\rm 2D, f} } (k_\parallel) $ is the effective two-dimensional polarization function, which only depends on $k_\parallel$ and takes the following form~\cite{Miransky:2015ava}:
\begin{eqnarray}
 \Pi_{ \mu \nu}^{ {\rm 2D, f} } (k_\parallel) \sim \frac{\, P_{\mu \nu}^{ E \parallel } \,}{\, (4\pi)^2 \,} 
  \left\{
 \begin{matrix}
 \frac{\, c k_\parallel^2 \,}{\, M_{ {\rm f} }^2 (B) \,} + \cdots, & |k_\parallel| \ll  M_{ {\rm f} } (B) \\
c' + \cdots ,& |k_\parallel| \gg  M_{ {\rm f} } (B)
 \end{matrix} \right.
\end{eqnarray}
with $1/(4\pi)^2$ typical for the loop integral. 
For $|k_\parallel| \gg M_{ {\rm f} }$, the result looks similar to that from the massless Schwinger model \cite{Schwinger:1962tp}. On the other hand, in the infrared limit $|k_\parallel| \rightarrow 0$, the quark mass cutoffs the infrared component so that $\Pi_{\mu \nu}^{ {\rm 2D, f} } (k_\parallel)$ vanishes, i.e., no screening mass is generated. But still the effective color-charge is strongly renormalized.

As the gauge field $A_\parallel$ is the ingredient of $E_\parallel$, at large $B$ and for {$k_\perp \ll M_{ {\rm f} }, B_{ {\rm f} }$}, the dielectric function approaches the infrared limit as
\be
\xi^B_\parallel ( k ) - 1 \, \sim \, c \, \frac{\, \lambda \,}{\, (4\pi)^2 \Nc \,} \frac{\, B_ { {\rm f} } \,}{\, M^2_{ {\rm f} } (B) \,} + \cdots,
\label{para}
\ee
where $\lambda = \Nc g^2$ is the 'tHooft coupling \cite{tHooft:1973alw} 
and $``\cdots"$ are terms from higher LLs of $O(B^0)$ according to the explicit $B$ dependences. Unless the effective mass $M^2_{ {\rm f} } (B)$ grows linearly with $B$, the dielectric function $\xi_\parallel^B$ can be large. Below we assume that the leading 
corrections $\lambda B_{ {\rm f} }/ ( \Nc M_{ {\rm f} }^2 )$ are small enough to avoid the $\beta$-function to change sign. In more realistic treatments with the renormalization group (RG) improvement, the vacuum coupling $g$ is replaced by a smaller screened coupling $g_B$, 
which is then frozen at the fixed point before the $\beta$-function flips the sign.  
In general, we expect that our one-loop calculations overestimate the magnitude of $\xi_\parallel^B$. On the other hand, the inclusion of higher order loop effects will not modify the basic structure of Eq.(\ref{para}) except the coefficient of $B_{ {\rm f} }/( \Nc M_{ {\rm f} } )$. For this reason we take the one-loop results as the representative of the leading $1/\Nc$ corrections.

The color-electric fields $E_\perp$ contain $A_0$ and $A_\perp$, and the corresponding polarization function is $\Pi_{ {\rm mix} }$. This contribution inevitably involves higher LLs, since the spin-flipping process would excite quarks from the LLL to the higher LLs. 
The infrared cutoff of $\xi^B_\perp$ is not $M^2_{ {\rm f} } (B)$ anymore, rather, $M^2_{ {\rm f} } (B) + 2nB_{ {\rm f} }$ with $n\ge1$. So we have in the infrared limit,
%
\be
\xi^B_\perp ( k ) - 1 \, \sim \, \frac{\lambda}{\, \Nc \,} \left( \sum_{n=1}\frac{\, c B_ { {\rm f} }/(4\pi)^2 \,}{\, M^2_{ {\rm f} } (B) + 2nB_{ {\rm f} } \,} - \Pi_{ {\rm q} }^{ B=0} \right)\,.
\label{perp}
\ee
Actually, due to the absence of LLL, the contributions with finite magnetic fields are smaller hence $\xi_\perp^B<1$. 
In this case, the low energy processes with effective transverse momentum $p_\bot \ll B_{ {\rm f} }$ are excluded in the first term, but not in the second one. As the effect of higher LLs is usually suppressed due to the larger transverse kinetic energy compared to LLL, we expect that $\xi^B_\perp ( k )$ changes more mildly than $\xi^B_\parallel ( k )$ with either $k$ or $B$.

After these qualitative analyses for very large magnetic fields, we now switch to the full one-loop expressions including all LLs, which allows us to study the transient regime from small to large magnetic fields. The expressions are ($i = \, \parallel, \perp$) \cite{Tsai:1974ap,Urrutia:1977xb}
\begin{widetext}
\begin{eqnarray}\label{xi0}
\xi_{i}^B (k)
&=&1 + { \lambda \over \, 2(4\pi)^2\Nc \,} 
 \sum_{{\rm f}=u,d,s}
\int_0^\infty\! {ds\over s}
\int_{-1}^{1}\!du \Big[\, {\rm e}^{ -s \left( M_{\rm f}^2(B) - {1 - u^2\over4} k_\parallel^2 + \phi ( s B_{ {\rm f} } ,u) k_\bot^2 \right) } N_{i} ( s B_{ {\rm f} }  ,u ) 
	- {\rm e}^{-s \left( M_{\rm f}^2 (0)-{1- u^2\over4} k^2 \right)} (1- u^2) \Big] \,, 
	\nonumber 
\\
&& \hspace{0.5cm}
N_\parallel (x,u)  = x \coth \left( x \right) \left( 1 - u^2 \right) \,,
\hspace{1cm}
N_\bot (x,u)  = {\, x \, \over \, \sinh \left( x \right) \,} \left[\, \cosh \left( x u \right) - u  \sinh \left( x u \right) \coth \left( x \right)\, \right] \,,
\end{eqnarray}
\end{widetext}
where $s$ and $u$ originate from the variable transformations of the proper-times involved in the two quark propagators of the loop. The difference between $\xi^B_\parallel$ and $\xi^B_\perp$ is summarized in the functions $N_{\parallel}$ and $N_{\perp}$, and
\be
 \phi (x,u) = {\, \cosh \left( x  \right) - \cosh \left( xu \right) \, \over \, 2 x \sinh \left( x \right) \,} \,.
\ee
In the UV limit ($x\rightarrow 0$), the asymptotic behaviors of these functions are
\be
N_i \rightarrow 1-u^2\,,~~~~~ \phi  \rightarrow \frac{\,1-u^2 \,}{4} \,,
\ee
with which one can ensure that the UV divergence from the domain $s \sim 0$ cancels in Eq.(\ref{xi0}). Meanwhile, in the IR limit ($x\rightarrow \infty $),
\be
N_{\parallel} \rightarrow x(1-u^2)\,,~~~N_{\perp} \rightarrow (1-u) x \, {\rm e}^{-(1-u)x} \,,~~~\phi \rightarrow \frac{1}{\, 2x \,} \,.
\ee
Using these expressions, one can recover the behavior of Eq.(\ref{para}) from the domain $s\sim \infty$. 

To evaluate the dielectric functions Eq.(\ref{xi0}), we consider the following schematic parameterization for the effective quark masses:
\be
M_{ {\rm f} } (B) = M_{ {\rm f} } (0) + C_B B_{ {\rm f} }^{1/2}
\label{eq:mass}
\ee
with $C_B (\geq 0)$ a common dimensionless parameter for all flavors. Which term dominates in Eq.(\ref{eq:mass}) depends on the strength of nonperturbative interactions \cite{Kojo:2014gha}. We take the effective quark masses in vacuum as $M_u(0) = 336~{\rm MeV}, M_d(0) = 340~{\rm MeV}$, and $M_s(0) = 486~{\rm MeV}$~\cite{David2008}.

\begin{figure}[!htb]
	\begin{center}
		\includegraphics[width=8cm]{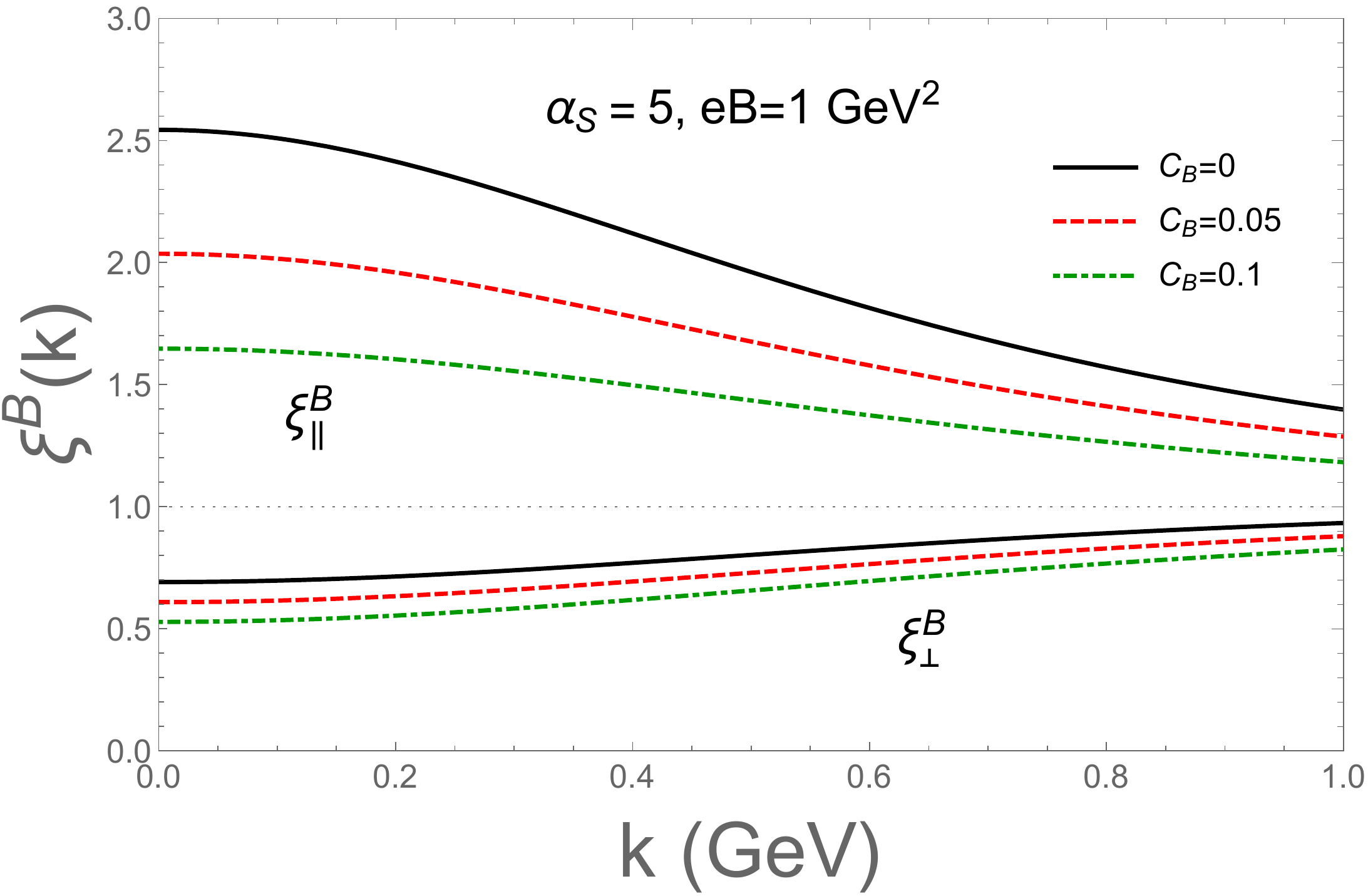}
		\caption{The dielectric functions $\xi_\parallel^B (k=k_3)~(>1)$ and $\xi_\bot^B (k=k_\bot)~(<1)$ as functions of the corresponding momentum $k$ for different coefficients $C_B=0,0.05$ and $0.1$. 
		The coupling constant in vacuum and the magnetic field are chosen to be $\alpha_S=5$ and $eB=1 \,{\rm GeV}^2$, respectively.} \label{fig:xi}
	\end{center}
\end{figure}

Shown in Fig.\ref{fig:xi} are the dielectric functions $\xi_i^B$'s as functions of the corresponding momenta: For the $\xi_{\parallel}^B (k)$ ($\xi_{\perp}^B (k)$), we take $k=k_3$ ($k=k_\perp$) by setting $k_0=k_\perp=0$ ($k_\parallel=0$). The vacuum coupling constant $\alpha_S = {g_0^2/4\pi}=5$ and the magnetic field $eB=1 \,{\rm GeV}^2$ are chosen for illustration.
The dielectric functions are sensitive to magnetic field and mass parameter in the infrared region $k\sim 0$; meanwhile at large momenta $k\sim 1 \,{\rm GeV}$, they approach the same vacuum value as expected. 
The numerical calculations also confirm that our qualitative analyses, according to the presence (absence) of LLL in Eq.(\ref{para}) [(\ref{perp})], capture the trends of the results employing all Landau levels. 
%

\subsection{Anisotropic confinement}

On the lattice, the string tension has been analytically computed in the strong coupling expansion \cite{Wilson:1974sk}. The leading order formula is given by \cite{Creutz1983}
\begin{eqnarray} \label{bare}
\sigma_0
= a^{-2} \ln \big[ N_{ {\rm c} } g^2(a)  \big] \,,
\end{eqnarray}
where $\sigma_0$ is the string tension in vacuum, $a$ is the lattice spacing, and $g(a)$ is the corresponding coupling. Since the expression is valid only for large coupling, we must regard the scale $a^{-1}$ to be sufficiently infrared; in other words, we must start with the action in which the RG evolution has been carried out to low enough energy.

Inspired by this formula, we estimate the ratios of the string tensions with and without magnetic fields as
\begin{eqnarray}\label{sigmaR}
\frac{\, \sigma_i^B \,}{\sigma_0} 
= \frac{\, \ln \left( \lambda / \bar{\xi}_{i}^B \right) \,}{  \ln \lambda  }
= 1 - \frac{\, \ln \bar{\xi}_i^B \,}{\, \ln \lambda \,} \,.
\end{eqnarray}
Here, we take $\bar{\xi}^B_i$ to be a constant, $\xi^B_i (k\rightarrow 0)$, as only infrared dynamics is important for confinement. The modifications of the string tensions originate entirely from the effective couplings $g^2/ \bar{\xi}_i^B$. Actually, the dielectric functions $\bar{\xi}_i^B$ affect the string tensions only logarithmically, which tempers the modifications at large $B$. 
Meanwhile, the logarithm at small $B$ behaves as $\ln (1+g^2 \Delta \Pi_{ {\rm q} }^B) \simeq g^2 \Delta \Pi_{ {\rm q} }^B$. 

Our treatment would have a solider theoretical ground in the large $N_c$ limit: $1 \ll \lambda \ll \Nc$, for which the ratio behaves as
\be
\frac{\, \sigma_i^B \,}{\sigma_0}  ~\rightarrow ~ 1 - \frac{\lambda}{\, \Nc \ln \lambda \,} \left( \Delta \Pi_{ {\rm q} }^B \right)_i  \,.\label{sigmaEp}
\ee
The $1/\Nc$ counting allows us to evaluate the string tension $\sigma$ (leading order in $\Nc$) and fermion loop effects $\bar{\xi}_i^B$ ($1/\Nc$-corrections) differently in the context of coupling $g$ expansion: the former requires the strong coupling picture while the latter is based on quasi-particle picture. These seemingly contradicting treatments are self-consistent if the quark loop corrections, of order $\lambda/\Nc$, can be regarded as small.
\begin{figure}[t]
	\begin{center}
	\vspace{0.5cm}
		\includegraphics[width=7.9cm]{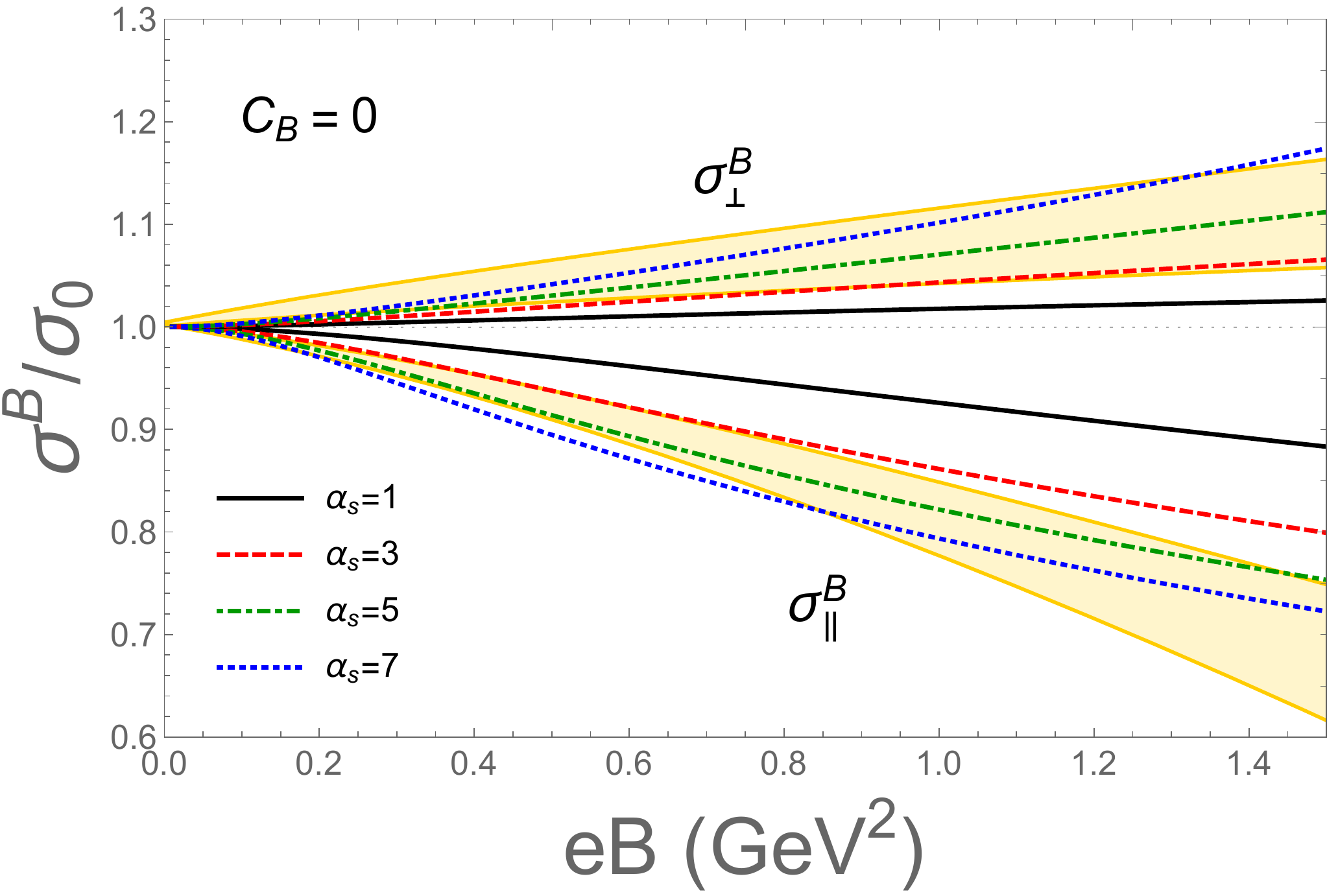}
		\caption{The longitudinal and transversal string tensions $\sigma_\parallel^B$ and $\sigma_\bot^B$ as functions of magnetic field $B$. We take constant quark masses by setting $C_B=0$ and the coupling constant is varied from $\alpha_S=1$ to $7$. The shadowed yellow bands are the continuum limits extrapolated from the lattice data~\cite{Bonati:2016kxj}.}
		\label{sigma_alpha}
	\end{center}
\end{figure}
\begin{figure}[t]
	\begin{center}
		\vspace{0.5cm}
		\includegraphics[width=8.5cm]{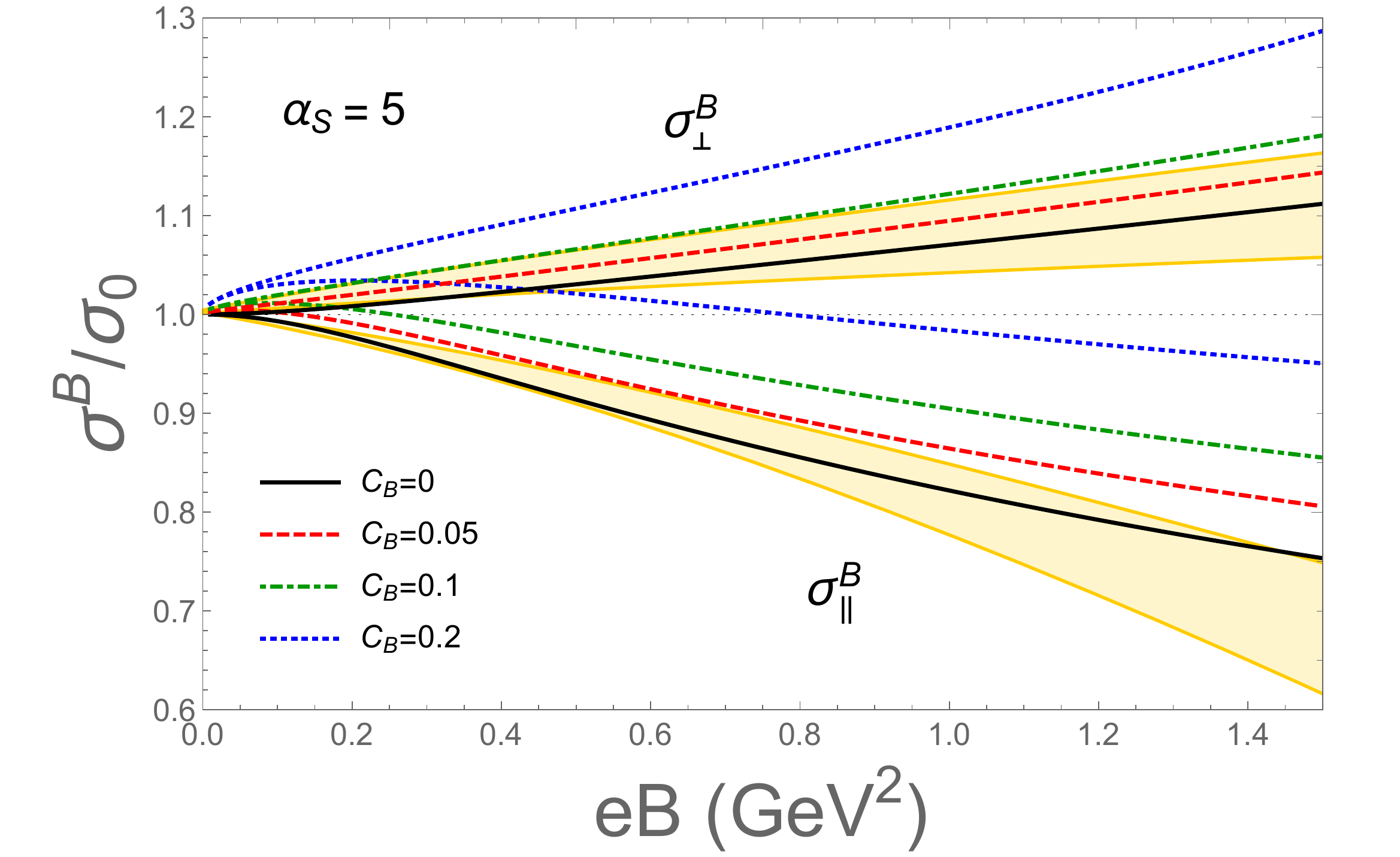}
		\caption{The same plot as Fig.\ref{sigma_alpha} except that we fix the coupling $\alpha_S=5$ and change the mass parameter $C_B$ from $0$ to $0.2$.}\label{sigma_CB}
	\end{center}
\end{figure}

Now we fix the number of color $\Nc=3$ and compare our results for $\sigma_\parallel^B$ and $\sigma_\bot^B$ based on the formula Eq.(\ref{sigmaR}) with those from lattice QCD~\cite{Bonati:2016kxj}. We constrain ourselves to the range $0\leq eB\leq1.5~{\rm GeV}^2$ beyond which the lattice results indicate that the string breaking effect becomes significant and the range of linear potential part is short 
-- in such domain, the concept of string tension seems somewhat obscure to us. Another reason for choosing this range is that our expansion in 
Eq.(\ref{sigmaEp}) 
would become invalid at too large $B$ such that $\lambda B/\Nc \gtrsim 1$.

Shown in Fig.\ref{sigma_alpha} are the string tensions as functions of $B$ with fixed mass parameter, $M_{\rm f} (B)=M_{\rm f} (0)$. As one of the biggest uncertainties in our analyses is the choice of the coupling constant $\alpha_S$, we vary it in a wide range from $1$ to $7$, refer to the review~\cite{Deur:2016tte}. The qualitative trends of the lattice results, $\sigma_\parallel^B <\sigma_0 < \sigma_\perp^B$, are captured for all couplings, and the quantitative agreements can be found for $\alpha_S=3,5$, and $7$. Next, shown in Fig.\ref{sigma_CB} are the same quantities as Fig.\ref{sigma_alpha}, but now we fix $\alpha_S=5$ and change the parameter $C_B$ from $0$ to $0.2$. 
For large $C_B~(\gtrsim 0.05)$, the $B$-dependent terms $C_B B_{\rm f}^{1/2}$ of the quark masses over-suppress the screening effects from quarks in the weak $B$ region, which renders the string tensions incompatible with the lattice results. 

\section{Summary and discussions}

In this letter, we have developed a simple framework to estimate the magnetic field dependence of the confinement dynamics. We estimate the dielectric functions based on the quasi-particle picture, which is valid at large $\Nc$, and take the one-loop results as the representative of such contributions. The dielectric functions enter the formulas for string tensions obtained in the strong coupling expansion and appear in the logarithmic forms. Numerical calculations at $\Nc=3$ of our formula seem to capture the tendency of the lattice results, provided that the effective quark masses do not radically grow with the magnetic fields.

Our framework combines the quasi-particle treatments for quarks with the collective dynamics of gluons. A framework of this sort may have a variety of applications, such as the system with external electric fields, rotational systems, and matter at high baryon density. We hope that further sophistication of the present framework will enable quantitative understandings of such systems.

\sect{Acknowledgments}
We appreciate C. Bonati's generosity for providing us their lattice data as shown in Figs.~\ref{sigma_alpha} and \ref{sigma_CB} 
and X.G. Huang for discussions. 
GC is supported by the National Natural Science Foundation of China (NSFC) with Grant No. 11805290 and Young Teachers Training Program of Sun Yat-Sen University with Grant No. 19lgpy282. TK is supported by the NSFC grant 11875144.

\end{document}